\documentclass [12pt,a4paper]{article}
\usepackage{psfig}
\usepackage{amssymb}

\def\slantfrac#1#2{\hbox{$\,^{#1}\!/_{#2}$}}
\def\mdot{{\raisebox{1pt}{\hbox{$\stackrel{\bullet}{M}$}}}\ \!\!}
\def\apgt{\ {\raise-.5ex\hbox{$\buildrel>\over{\scriptstyle\sim}$}}\ }
\def\aplt{\ {\raise-.5ex\hbox{$\buildrel<\over{\scriptstyle\sim}$}}\ }

\def\dfrac#1#2{{\displaystyle\frac{#1}{#2}}}
\def\q{\linebreak}

\begin{document}

\title{A Model of Superoutbursts in Binaries of SU~UMa Type}

\author{D.V.Bisikalo$^1$, A.A.Boyarchuk$^1$,\\
P.V.Kaygorodov$^1$, O.A.Kuznetsov$^{1,2}$ and T.Matsuda$^3$\\[5mm]
$^1$ Institute of Astronomy RAS, Moscow, Russia\\
$^2$ Keldysh Institute of Applied Mathematics, Moscow, Russia\\
$^3$ Department of Earth and Planetary Sciences, Kobe
University,\\ Kobe, Japan\\}

\date{}
\maketitle

\begin{abstract}

A new mechanism explaining superoutbursts in binaries of \q SU~UMa
type is proposed. In the framework of this mechanism the accretion
rate increase leading to the superoutburst is associated with
formation of a spiral wave of a new ``precessional" type in inner
gasdynamically unperturbed parts of the accretion disc. The
possibility of existence of this type of waves was suggested in
[\ref{prec_1}]. The features of the ``precessional" spiral wave
allow explaining both the energy release during the outburst and
all its observational manifestations. The distinctive
characteristic of a superoutburst in a SU UMa type star is the
appearance of the superhump on the light curve. The proposed model
reproduces well the formation of the superhump as well as its
observational features, such as the period that is 3--7\% longer
than the orbital one and the detectability of superhumps
regardless of the binary inclination.

\end{abstract}

\section{Introduction}

Binaries of SU~UMa type are dwarf novae with orbital period
shorter than 3 hr that display superoutbursts. At present, this
traditional definition is expanded to include the requirement of
presence of the superhump during the superoutburst
[\ref{warner}]. Normal outbursts in binaries of this type are
rather short and irregular. These ones are explained well by
standard models of dwarf novae (see, e.g., [\ref{warner}]).
Superoutbursts are essentially more prolonged, more
rare but more regular. For instance, binary OY~Car
displays normal outbursts each 25--50 days with amplitude $\sim
3^m$ and duration about 3 days. Superoutbursts are repeated with
period $\sim300$ days, their amplitude reaches $4^m$ and the
duration is as large as 2 weeks. The analysis of observational
data shows that the majority of superoutbursts has practically
identical profile: the brightness rises sharply for a time of
$\sim0.1$ of superoutburst duration, then the brightness
diminishes slowly during a time of $\sim0.8$ of superoutburst
duration (an extend sloping ``plateau"), after that the system
rapidly comes back to the quiescent state. Superhumps with
period $P_s$ of a few per cents longer than the orbital period
$P_{orb}$ have been observed in every SU~UMa star for which high
speed photometry during a superoutburst has been obtained. At
their full development the superhumps have a range of $\sim
0.3-0.4^m$ and are equally prominent in all SU~UMa stars
independent of inclination.

The presence of periodic superoutbursts accompanied by superhumps
place dwarf novae of SU~UMa type among the most enigmatic
phenomenon in astronomy. Despite plenty of both
observational data and theoretical models, the
understanding of its nature is far from completion. In present
work we suggest a new mechanism for explanation of
superoutbursts in SU~UMa stars. The essence of this mechanism is
as follows: (i) during the time between the superoutbursts the
accretion disc is formed, it accumulates matter and becomes more
dense as compared to the matter of the stream, a gasdynamically
non-perturbed zone is formed in the inner part of the disc; (ii)
in accordance with [\ref{prec_1}], a spiral wave of
``precessional" type is generated in the disc's inner part were
gasdynamical perturbations are negligible; (iii) the formation
of the wave is accompanied by substantial (up to the order of
magnitude) increase of the accretion rate; (iv) retrograde
precession of the spiral wave as well as the compactness of
inner zone with increasing energy release explain the superhump
period which is longer than orbital one as well as the superhump
being observable independently of the inclination of the binary.
Note, the assumption on the generation of gasdynamically non-perturbed
region as far as the matter accumulates in the disc up to some
limit permit us to stay in a paradigm of constant mass transfer
rate in the binary system. This substantially simplifies the
model, since one doesn't need any more to appeal to variations
of conditions in mass-losing star for the explanation of
periodic superoutbursts anymore.

The presentation of the new mechanism explaining the
superoutbursts in SU~UMa stars is given in the paper as follows:
Section~2 contains a brief review of observational data and
theoretical models for superoutbursts; Section~3 contains the
results of 3D gasdynamical simulation of the morphology of
gaseous flows in semidetached binaries as well as the
description of basic features of the ``precessional" spiral wave
forming in the inner part of the disc; the comparison of
observational manifestations of superoutbursts and computational
results confirming the consistency of the new model is given in
Section~4.

\section{Observations and theoretical models\\ for superoutbursts}

Following to Warner (see monograph [\ref{warner}]), let us
summarize the basic features of superoutburst and superhump on
SU~UMa variables.

\bigskip

The superoutburst peculiarities can be stated as:

\begin{enumerate}

\item
it is $\sim 0.5-1^m$ brighter than a normal outburst; total energy
released in a superoutburst is $E \simeq 10^{40}$~erg, so it is
$\sim$10 times larger than that for a normal outburst;

\item
its duration is few times larger than that for normal outburst
and can be several weeks long;

\item
the intervals between superoutburst are rather long and, as
a rule, are hundreds of days, but for some systems can
reach few thousands of days; the intervals are approximately
constant;

\item
all superoutbursts have the same profile: rapid rise, extended
sloping plateau, rapid decline;

\item
the slope of the plateau is almost invariant and equal to $\sim 9
\pm 1$~day/mag, i.e. the brightness diminishes one stellar
magnitude during approximately 9 days; since the duration of
superoutbursts is ranged from $\sim$10~to $\sim$30~days, the
changing of brightness in plateau is $\sim 1-3^m$;

\item
the brightness and color changes in the rise to a superoutburst
are in general indistinguishable from those for a normal outburst
in the same system, so a superoutburst begins to develop as a
normal outburst;

\item
there are no recorded instance of a normal outburst occurring
during a superoutburst, or immediately at the end of it; basing on
this fact it is usually concluded that normal outbursts and
superoutbursts are not independent phenomena;

\item
the superhumps appear some time after maximum of
superoutburst (the so called interregnum, this time is, as a
rule, from $\slantfrac1{20}$ to $\slantfrac12$ of the plateau
length);

\item
in the eclipsing SU~UMa systems (e.g., OY~Car and Z~Cha) light
curves have recurrent dips at phases $\sim$0.25 and 0.75, these
ones are usually ascribed to increases in vertical thickness of
the disc at the corresponding phases; UV flux distributions also
show minima at phases $\sim$0.2 and~0.8;

\item
asymmetrical line profiles show that during superoutbursts a
substantial fraction of the disc is in non-circular motion; the
non-circular component rotates  with period
$P_s$ not $P_{orb}$, so it is intimately linked with the
superhump process;

\item
for some SU~UMa stars (e.g., VW Hyi) after the rapid brightness
decrease at the end of plateau there are both normal orbital
humps and also a modulation at the superhump period shifted in
phase by $\sim 180^\circ$; these are known as the ``late
superhumps";

\end{enumerate}

The superhump peculiarities can be stated as:

\begin{enumerate}

\item[12.]
superhumps have been observed for every SU~UMa star for which high
speed photometry during a superoutburst has been obtained;

\item[13.]
as mentioned above, the asymmetrical line profiles show that the
superhump phenomenon is linked with non-circular motion of the
disc components;

\item[14.]
the superhump period is 3--7\% longer than the orbital period;

\item[15.]
for some cases the substantial decrease (up to 1.25\%) of the
superhump period during the superoutburst is observed;

\item[16.]
superhumps are equally prominent in all SU~UMa stars
independent of inclination (even in stars like V436 Cen, WX
Hyi and SU UMa, which have no detectable orbital humps during
quiescence);

\item[17.]
at their full development the superhumps have a range of $\sim
0.3-0.4^m$;

\item[18.]
usually the amplitude of superhumps decreases faster than the
system brightness, causing them to disappear before the end of the
extended slope plateau of the superoutburst;

\item[19.]
multicolor photometry of superhumps shows that the superhump
light is bluest at minimum and reddest at superhump maximum,
thus there appear to be an inverse correlation between color
temperature and brightness of the superhump;

\item[20.]
the eclipse mapping technique permits to reveal the superhump
light sources, it was found that the main sources are located in
three regions in the disc.

\end{enumerate}

A number of models were suggested to explain the peculiarities
of superoutbursts and superhumps enumerated above. The first
models of superoutbursts and superhumps were based on possible
non-synchronously rotating mass-losing star
[\ref{vogt74}], intermediate polar model [\ref{papa78}],
slight eccentricity of binary orbit [\ref{papa79}],
mass transfer variations [\ref{vogt80},\ref{osaki85}],
disc instability [\ref{osaki89}], as well as combination of two
latter [\ref{duschl89}]. The review of early models as well as
its criticism can be found in the monograph by Warner
[\ref{warner}].

In 1982 Vogt [\ref{vogt82}] suggested that the accretion disc
takes up an elliptical shape during superoutburst, later
Osaki~[\ref{osaki85},\ref{osaki89}] and Mineshigi [\ref{mine88}]
proposed that a slowly precessing elliptical disc develops during
superoutburst.  Whitehurst [\ref{white88},\ref{white91}] confirmed
numerically the possibility of disc's precession for sufficiently
small values of $q$.  This model currently is thought to be the
essence of the process producing superoutbursts in SU~UMa stars.
The precession appearance in this model is due to the disc
instability caused by Lindblad eccentrical resonance 3:1
[\ref{lubow91}]. This resonance\footnote{Nonaxisymmetric modes in
the disc are expressed as $\exp[i(k\varphi-l\Omega t)]$ and
$(k,l)$ are an integer pair specifying a particular mode. Lubow
[\ref{lubow91}] have shown that the density perturbations produced
by tidal perturbation with (3,3) couples with an imposed eccentric
perturbations with (1,0) to excite two-armed spiral density waves
with (2,3) at the eccentric inner Lindblad resonance which is
given by $\omega=m\Omega/(m-2)$ for $m=3$, that is the 3:1
resonance.} is inside the disc only for sufficiently small values
of mass ratio $q\aplt0.22$ (the radius of the disc can be
determined, for instance, using the Paczy\'nski formula
[\ref{pac77}]), the latter is in a good agreement with the typical
values of $q$ for SU~UMa stars. Thus in accordance with
[\ref{osaki89},\ref{lubow91},\ref{osaki96}], the superoutburst
onset can be explained as follows: in the coarse of accretion the
matter is accumulated in the disc so the disc radius increases;
the short normal outbursts occurs during this time, the latter are
caused by the same thermal instability (see, e.g.,
[\ref{a1},\ref{a2}]) as that of ordinary dwarf novae of U Gem
type; a tidal instability occurs only when a normal outburst
pushes the outer edge of the disc beyond a critical radius for the
eccentric 3:1 Lindblad resonance, that is, a normal outburst
triggers a superoutburst. During the superoutburst the shape of
disc becomes elliptical, the disc begins to precess, and the
superhump is formed.

Yet this model can not explain {\it all} observational
manifestations of SU~UMa stars. Firstly, it can't explain the
appearance of superhump for binaries that don't display the
normal hump in quiescence, i.e. for {\it near face-on} SU~UMa
stars (e.g., V436~Cen, WX~Hyi, SU~UMa [\ref{warner}]).

The model of precessing accretion disc also can't explain the
appearance of superoutbursts and superhumps in SU~UMa binaries
with relatively {\it large mass ratio} ($q>0.22$, e.g., VY Scl
[\ref{murray00}]).

The ``standard" model also fails in explanation of {\it late
superhumps}. \q Osaki [\ref{osaki85}]
and Whitehurst [\ref{white88}]
proposed that the eccentric disc survives for several days after
the end of superoutburst, and the modulation of ``bright spot"
brightness produces late superhumps (see also [\ref{rolfe01},
\ref{ham}]). But an independent determination of disc eccentricity
for OY~Car doesn't meet this model
[\ref{woerd88},\ref{hessman92}].

Taking into account all these inconsistencies there are some
doubts in adequacy of existing models explaining superoutbursts
and superhumps in SU~UMa binaries.

\section{Results of 3D gasdynamical simulation of cool
accretion discs. The peculiarities of ``precessional" spiral
wave.}

Results of both qualitative analysis and 3D gasdynamical
simulations of the morphology of gaseous flows in semidetached
binaries when the gas temperature is low ($\sim 10^4$~K) permit us
to reveal the basic features of the structure of cool accretion
discs [\ref{prec_1},\ref{di12}]. In general, the flow structure is
qualitatively the same as for the case of high gas temperature
[\ref{di8}--\ref{di11}], namely: the gasdynamical structure of
gaseous flows is governed by the stream of matter from $L_1$,
accretion disc, circumdisc halo and circumbinary envelope; the
interaction between the stream and the disc is shock-free; the
interaction of matter of circumdisc halo and circumbinary envelope
with the stream results in the formation of the shock -- ``hot
line" -- located along the edge of the stream. At the same time
the gas temperature decrease leads to a number of differences.
Thus the cool accretion disc becomes sufficiently more dense as
compared to the matter of the stream, the disc is thinner and has
more circular form. The second arm of tidal spiral shock
(discovered in [\ref{Sawada86}--\ref{spiral2}]) is formed, the
both arms don't reach the accretor but are located in the outer
part of the disc. Taking into account that the stream acts on the
dense inner part of the disc weakly as well as that all shocks
(``hot line" and two arms of tidal wave) are located in the outer
part of the disc we can identify a new element of flow structure
for low-temperature case: the inner region of the accretion disc
where the impact of gasdynamical perturbations is negligible.

Formation of gasdynamically non-perturbed region in the
inner part of the disc allows to consider the latter as an slightly
elliptical disc with typical size of $\sim0.2-0.3A$ ($A$ is the
binary separation) embedded in the gravitational field of binary.
It is known (see, e.g., [\ref{warner},\ref{kumar}]), that the
influence of companion star results in precession of orbits of
particles rotating around of the binary's component. The
precession is retrograde and its period increases with approaching
the accretor. We have shown in [\ref{prec_1}] that the retrograde
precession with specified law of precession rate results in
formation of the density spiral wave of ``precessional" type in the
inner part of the disc. This wave is formed by apastrons of
flowlines and its appearance leads to growth of radial component
of matter flux $F_{rad} \propto \rho v_r$ due to increasing of
both density $\rho$ and radial velocity $v_r$.  The increasing of
radial component of matter flux after passing the wave results in
increasing of accretion rate in the region where ``precessional"
wave approaches the accretor.

Right panels of Fig.~1 depict the density distribution and
velocity vectors in the equatorial plane for four instants of
time (beginning from some chosen one and then more with the
interval of one orbital period), namely for $t=t_0$,
$t_0+P_{orb}$, $t_0+2P_{orb}$, $t_0+3P_{orb}$. Left panels show
the distribution of radial flux of matter in the inner parts of
the disc for the same instants of time. The analysis of these
results confirm that:  the dense accretion disc as well as the
compact circumdisc halo are formed; the interaction of matter of
circumdisc halo and circumbinary envelope with the stream
results in the formation of the shock -- ``hot line" -- located
along the edge of the stream; the two-armed tidal spiral shock
is formed, both its arms don't reach the accretor but are
located in the outer part of the disc.  We also can see one more
spiral wave in the inner part of the disc.

\renewcommand{\thefigure}{1}
\begin{figure}
\begin{center}
\hbox{\hspace*{2cm}\psfig{file=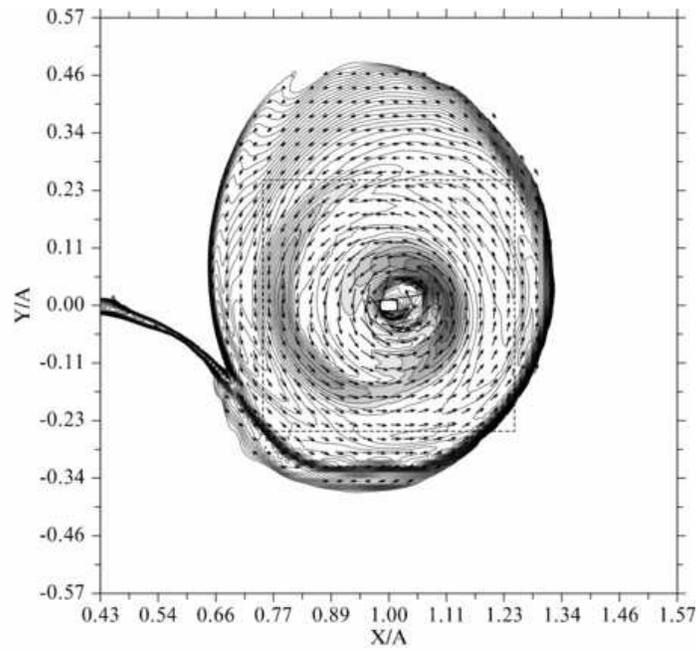,height=9cm}}
\vspace*{0.5cm}
\hbox{\hspace*{2cm}\psfig{file=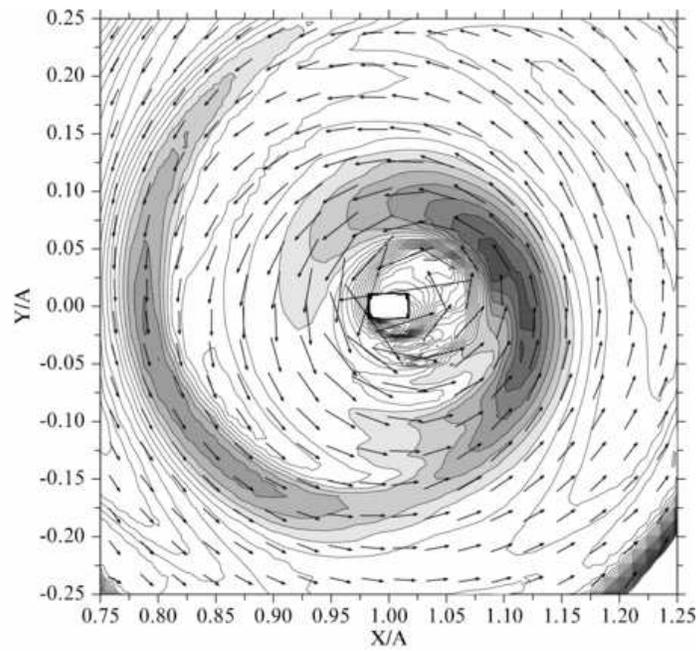,height=9cm}}
\vspace*{-1.5cm}
\end{center}
\caption{\small
{\it Left panel:} density isolines and velocity vectors in the
equatorial plane of binary; {\it right panel:} the same for the
central part of the disc.}
\end{figure}

\renewcommand{\thefigure}{1 {\it (continued)}}
\begin{figure}[p]
\begin{center}
\hbox{\hspace*{2cm}\psfig{file=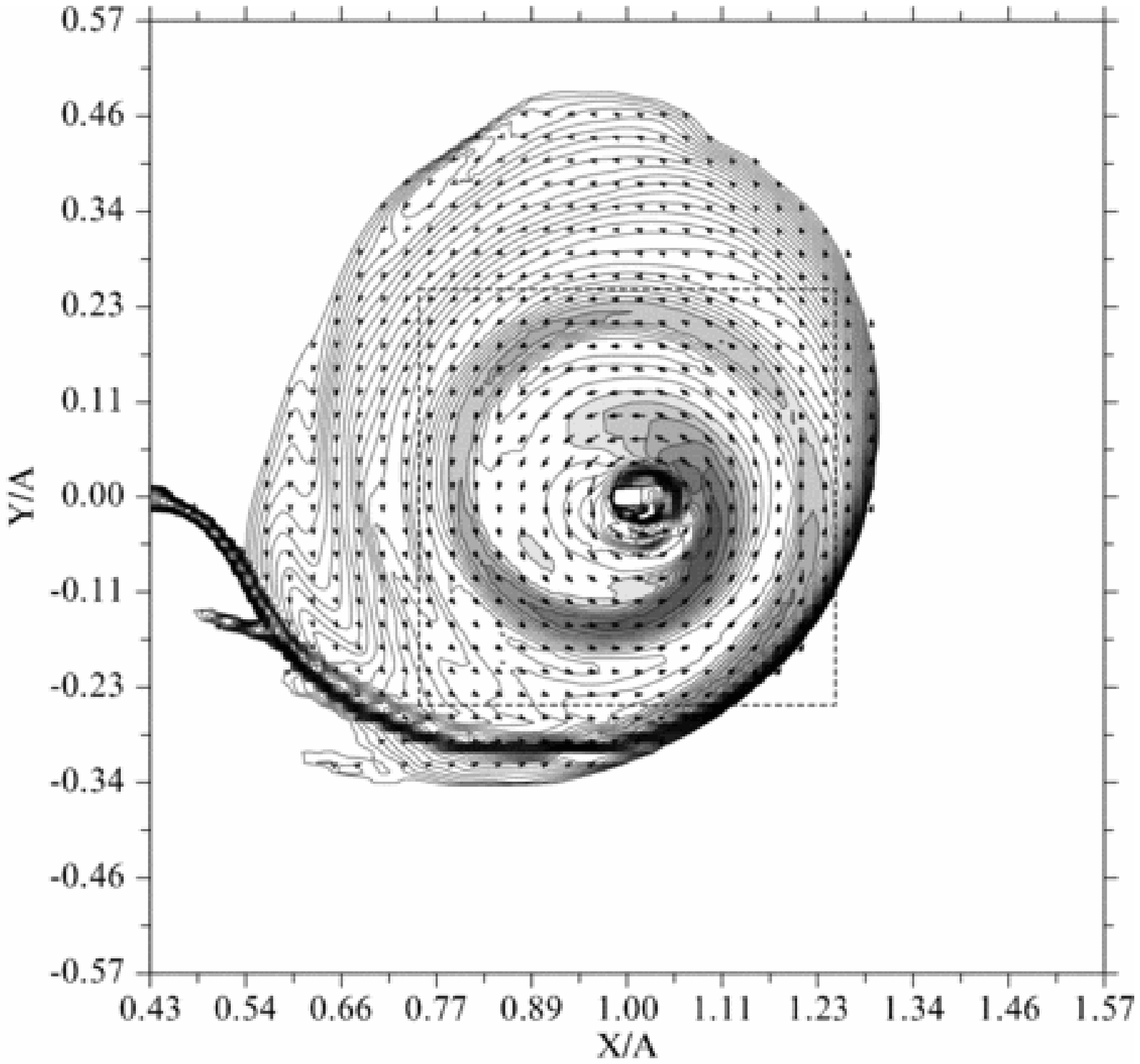,height=9cm}}
\vspace*{0.5cm}
\hbox{\hspace*{2cm}\psfig{file=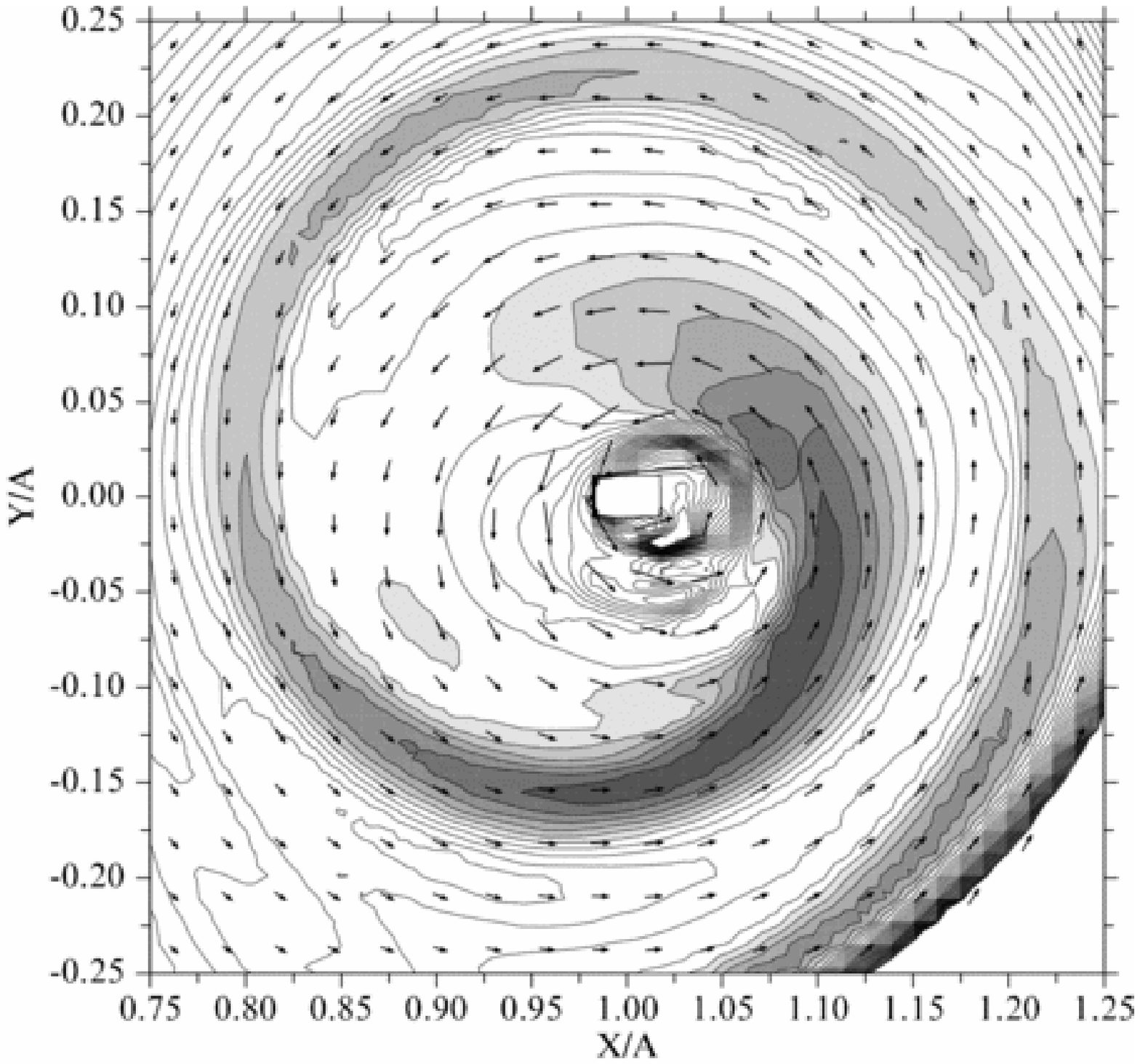,height=9cm}}
\vspace*{-1.5cm}
\end{center}
\caption{\small The same as in Fig.~1 but after
one orbital period.}
\end{figure}

\renewcommand{\thefigure}{1 {\it (continued)}}
\begin{figure}[p]
\begin{center}
\hbox{\hspace*{2cm}\psfig{file=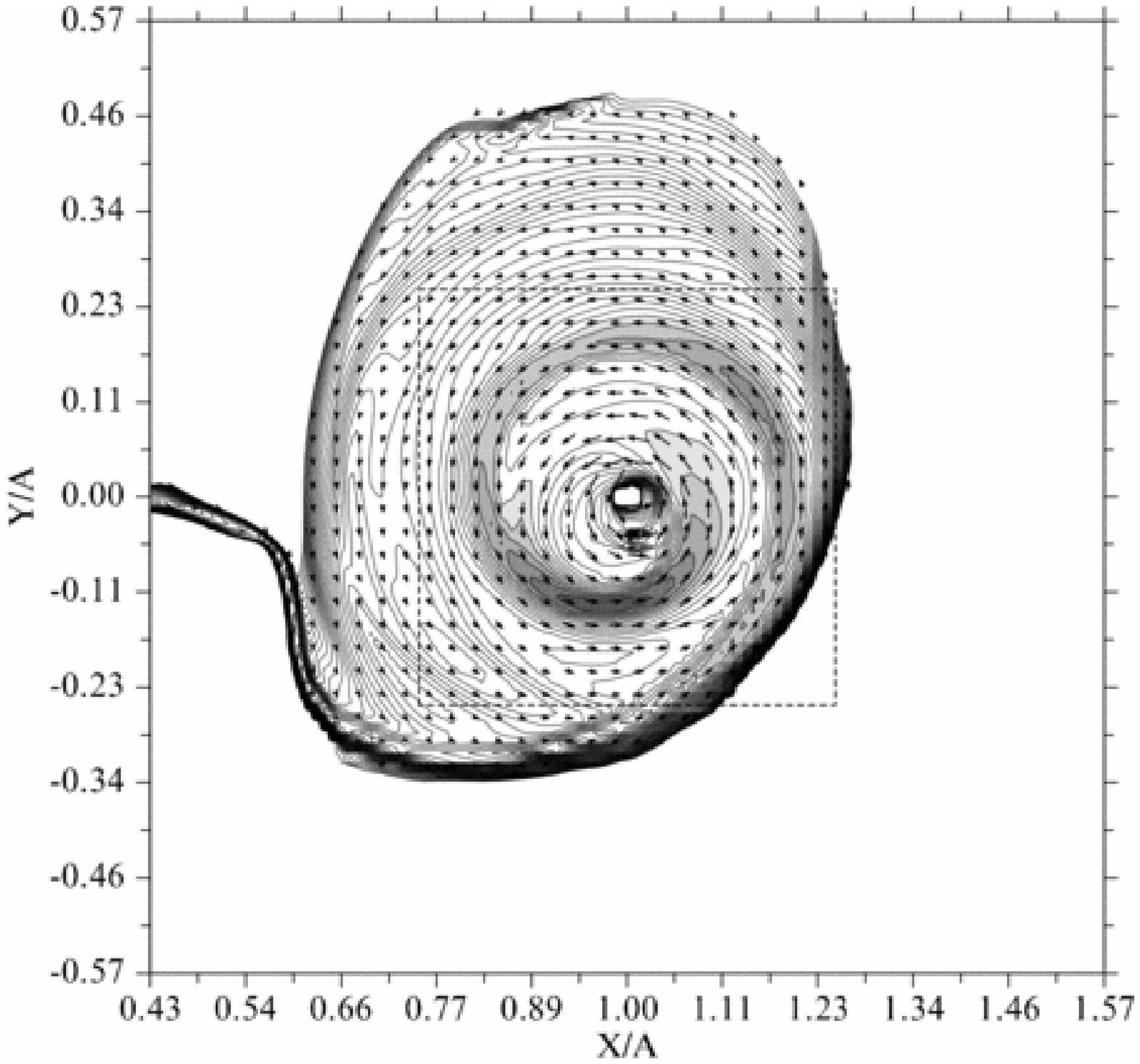,height=9cm}}
\vspace*{0.5cm}
\hbox{\hspace*{2cm}\psfig{file=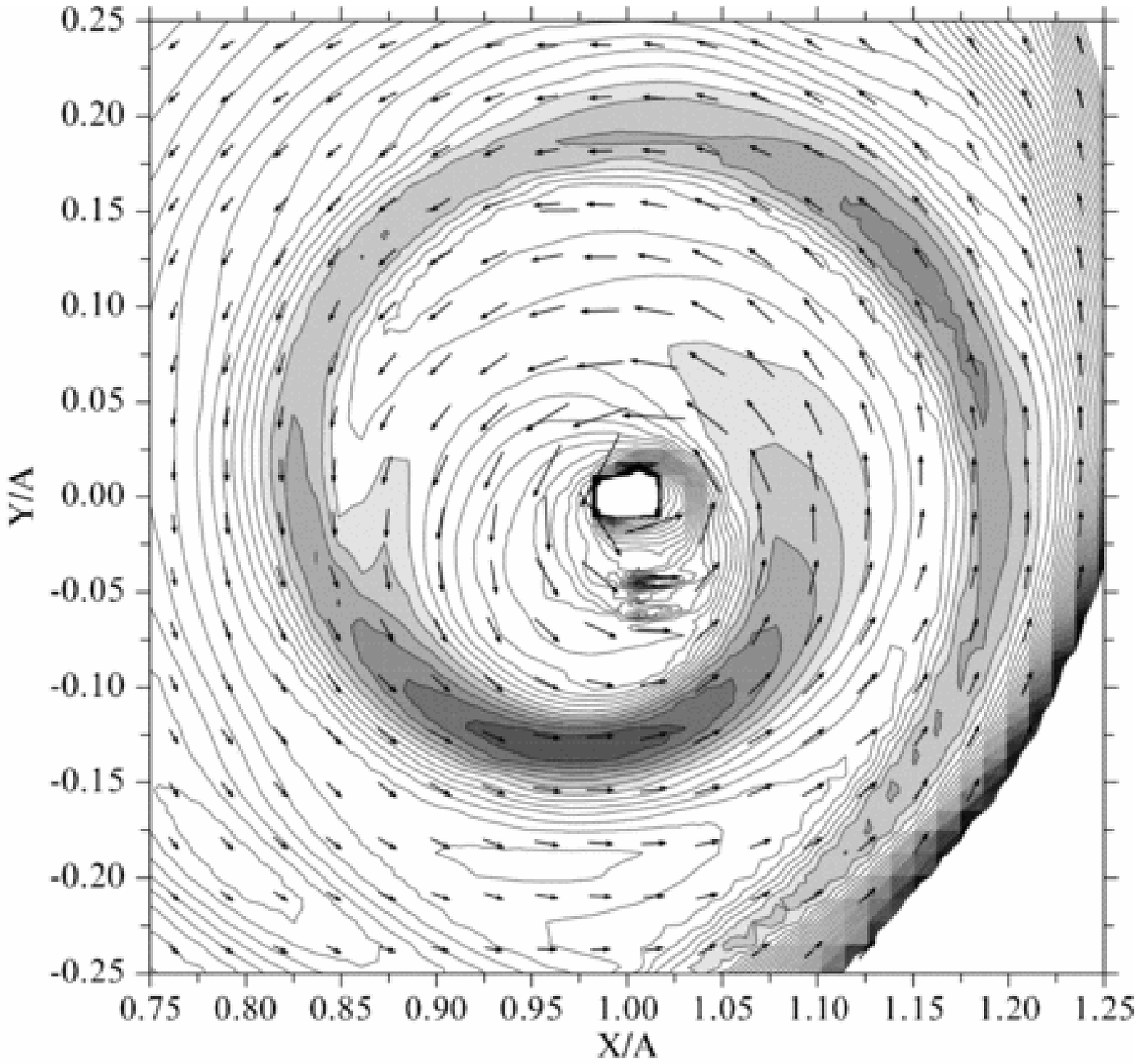,height=9cm}}
\vspace*{-1.5cm}
\end{center}
\caption{\small The same as in Fig.~1 but after
two orbital periods.}
\end{figure}

\renewcommand{\thefigure}{1 {\it (continued)}}
\begin{figure}[p]
\begin{center}
\hbox{\hspace*{2cm}\psfig{file=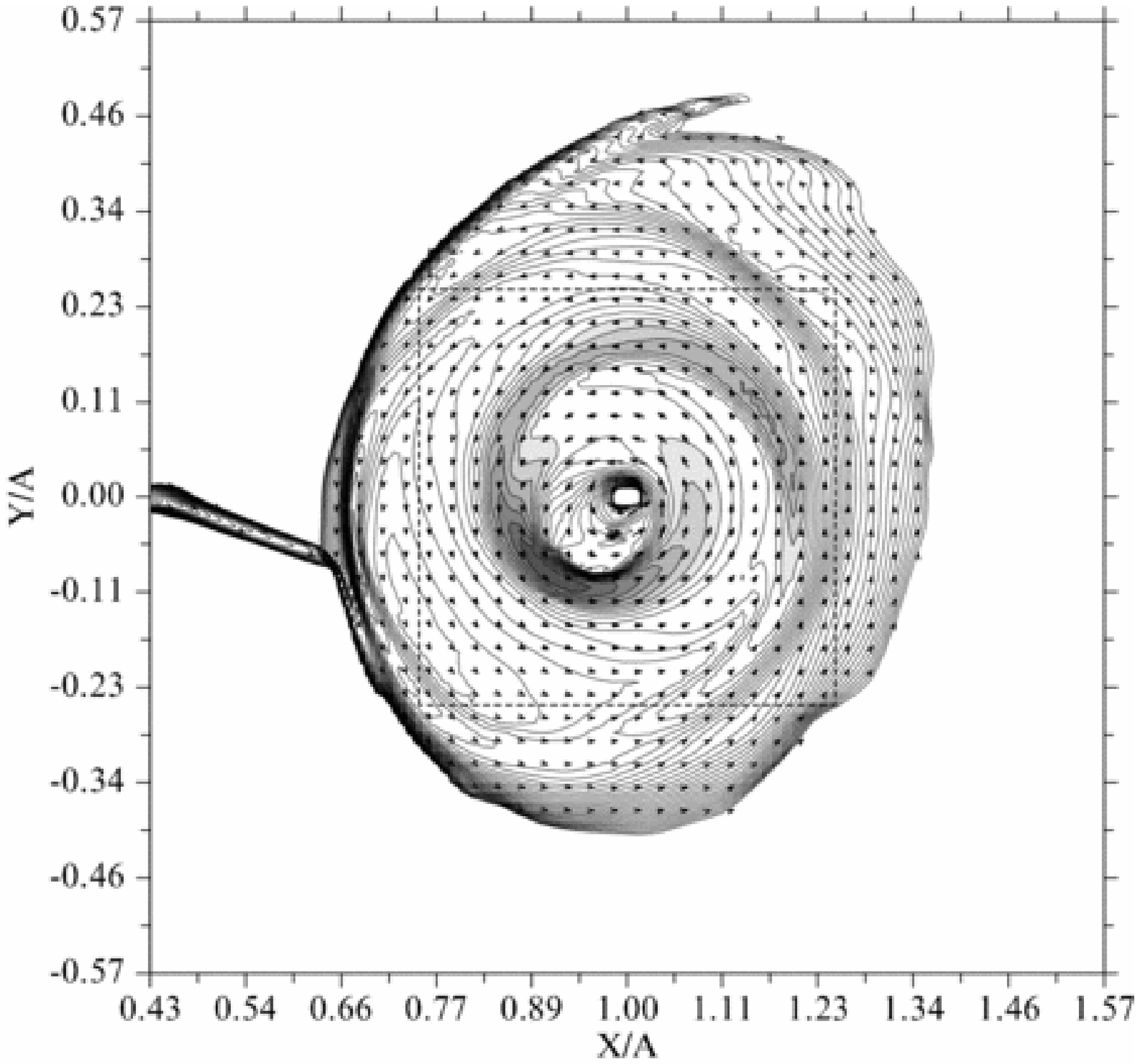,height=9cm}}
\vspace*{0.5cm}
\hbox{\hspace*{2cm}\psfig{file=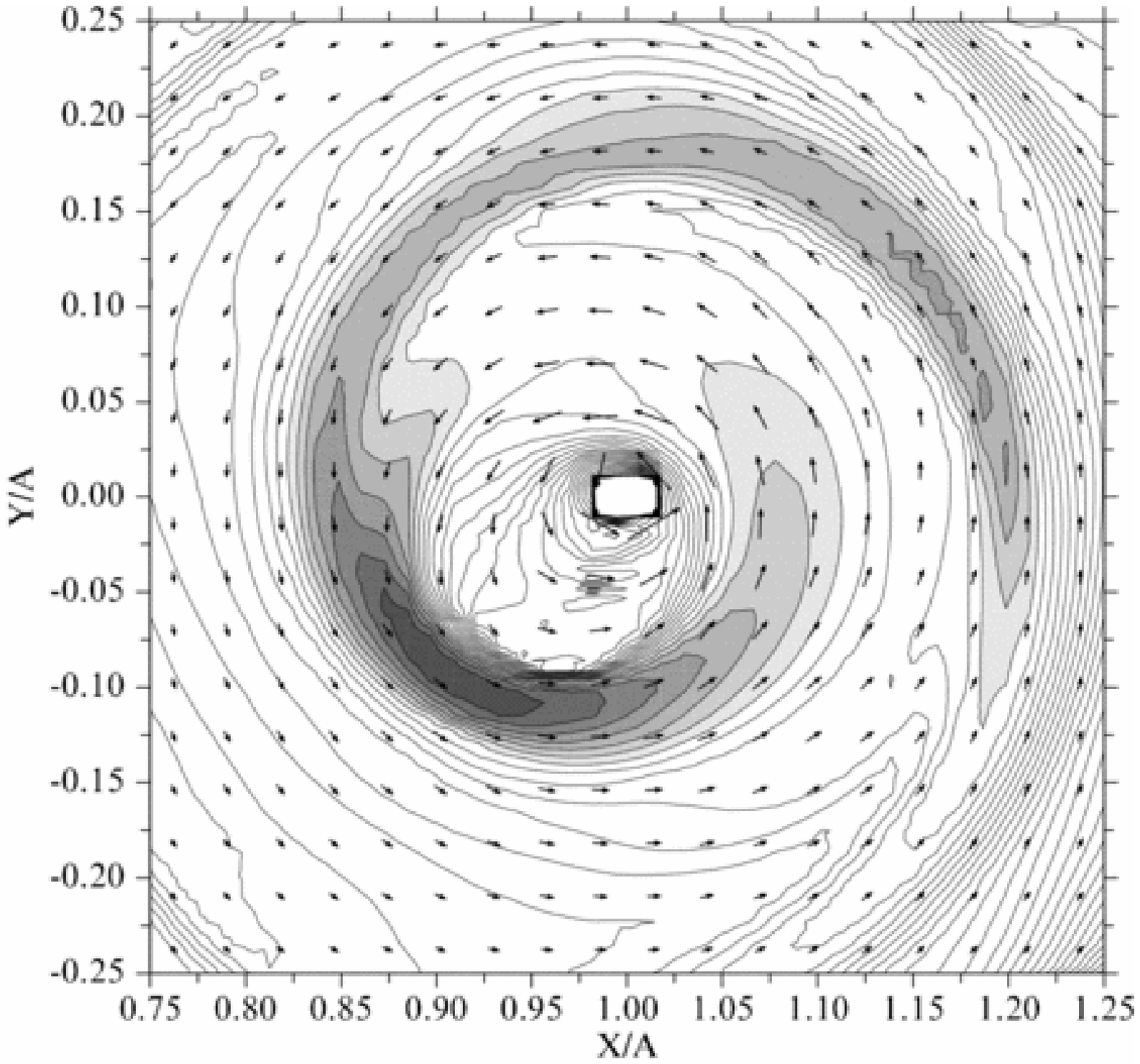,height=9cm}}
\vspace*{-1.5cm}
\end{center}
\caption{\small The same as in Fig.~1 but after
three orbital periods.}
\end{figure}

We have conducted our simulations for the binary with
characteristics of the dwarf novae IP Peg: $M_1=1.02M_\odot$,
$M_2=0.5M_\odot$, $A=1.42R_\odot$. Considering the computational
results for four instants of time we can see that the inner
spiral wave precesses retrogradely and the velocity of its
revolution in inertial frame (i.e. in the observer's frame)
equals $\approx -0.13$ revolution per one orbital period of
binary. Analyzing the distribution of radial flux of matter in
the disc we have found that it increases with approaching to the
accretor and in the maximum it is more than 10 times greater
than in other parts of the disc. Thus we can expect more than
tenfold increase of the accretion~rate.

Of course, the outburst of IP~Peg has characteristics that differ
from those of superoutbursts in SU~UMa binaries due to relative smallness
of the disc and, correspondingly, the smallness of the region
which isn't subjected to gasdynamical perturbations. Indeed, for a
typical binary of SU~UMa type the components' mass ratio is
$q=M_2/M_1 \approx 0.1$ and this value is substantially smaller than
that for IP~Peg. Decreasing of the components' mass ratio implies
that the size of Roche lobe increases ($x_{L_1}\to A$ as $q\to0$,
here $x_{L_1}$ is the distance between the accretor an the inner
Lagrangian point $L_1$), hence forming accretion disc should have
a larger size. In accordance to Paczy\'nski [\ref{pac77}], the
size of accretion disc depends on $q$ as
$$
\frac{R_d}{A}=\frac{0.6}{1+q}\,,$$
i.e. it can reach the value of
$0.54A$ for SU~UMa stars. Consequently, the region
without gasdynamical perturbations can be rather extended, so the
mechanism resulting in the formation ``precessional" could be
effective.

To be convinced in this, we have conducted gasdynamical
simulation for a binary with parameters of OY Car ($q$=0.147):
$M_1=0.95M_\odot$, $M_2=0.14M_\odot$, $A=0.69R_\odot$. Analysis
of the results of simulation shows that ``precessional" wave is
initiated at the distance $\approx0.25A$ from the accretor, the
velocity of its revolution turns out to be equal to $\approx
-0.03\div-0.04$ revolution per one orbital period of binary, and
the accretion rate after passing the wave increases more than 10
times.

Kinematical properties of the ``precessional" wave as well as an
substantial growth of the accretion rate due to this wave
permit us to engage it for the explanation of superoutbursts and
superhump in SU~UMa stars.

\section{Foundations of the new mechanism for\\
superoutburst.}

Let us review the basic statements of the suggested mechanism
explaining superoutbursts and superhumps in binaries of SU~UMa
type. In the framework of this mechanism the formation of
accretion disc is suggested, the accumulation of matter in the
disc makes it more dense as compared to matter of the stream
from $L_1$, hence the inner part of the disc turns out to be not
subjected by gasdynamical perturbations. A spiral density wave
of ``precessional" type is formed in the gasdynamically
non-perturbed parts of the disc; the formation of the wave is
accompanied by substantial (up to the order of magnitude)
increase of the accretion rate. The growth of accretion rate
results in brightening of the binary star, i.e.  developing of
superoutburst. Retrograde precession of the density wave with
the rate of $\sim$ few hundredth of revolution per one orbital
period of binary as well as the compactness of energy release
zone permits to explain the formation of superhump as well as its
observational peculiarities.

These observational peculiarities of superoutbursts and
superhumps in SU~UMa stars were enumerated in Section~2.
To be sure that the ``precessional" spiral wave can explain
these peculiarities let us provide a consequential, item by item
comparison between computed and observational peculiarities
of both superoutbursts and superhumps.


{\bf Energy release, recurrence time and duration of superoutburst
(peculiarities \#1--\#3 from Section~2).} Our mechanism suggests
that energy release and period of superhumps are determined by the
mass and accumulation time of the disc. During the superoutburst
approximately a half of the disc mass is accretted (see, e.g.,
[\ref{warner}]), this results in releasing of energy $E \simeq
10^{40}$ erg. Assuming that a half of binding energy is radiated
we can evaluate the mass of accreted part of the disc
($\slantfrac{1}{2}m_d$) using the formula $E \simeq
\slantfrac12\dfrac{G M_1}{R_1}\cdot\slantfrac12m_d$. Adopting
characteristic parameters for SU~UMa stars $M_1 \simeq 1 M_\odot$
and $R_1 \simeq 10^9$см, we get $m_d \simeq 1.5~10^{-10} M_\odot$.
Since the outburst recurrence time (i.e. the time of accumulation
of mass $\slantfrac12m_d$) is approximately equal to one year we
can evaluate the value of mass transfer rate as
$10^{-10}M_\odot$/year, the latter is in a good agreement with the
standard evaluation of mass transfer rate for cataclysmic
variables. Variation of recurrence time for different binaries
can be easily explained by variations of value of mass transfer
rate, range of the latter being reasonable --
for recurrence time $\sim100$~days mass transfer rate is
$\mdot\simeq3\cdot10^{-10}M_\odot$/year and
for recurrence time $\sim1000$~days it is
$\mdot\simeq3\cdot10^{-11}M_\odot$/year.
A strong regularity of
superoutbursts is determined by the constant rate of mass transfer
(we can neglect the evolutionary changes on these timescales).
Duration of superoutburst is determined by the ratio of the mass
of accretted matter to the accretion rate, the latter is
approximately equal to the mass transfer rate during quiescence so
increasing of accretion rate up to the order of magnitude (or even
more) during superoutburst gives the ratio of recurrence time to
the superoutburst duration as 10--20, that is in a good agreement
\q with observations as well.


{\bf The superoutburst profile (peculiarities \#4 and \#5 from
Section~2).} The rapid brightening at the superoutburst onset is
stipulated by the appearance of ``precessional" spiral wave in the
disc and induced increase of accretion rate, the latter is the
result of the more effective outward transport of angular momentum
in the disc. In the quasi-steady mode the angular momentum
transfer rate is apparently constant. In the framework of proposed
mechanism the inner regions of the disc are accreted first, and
after that the more outer regions. Due to the law of angular
momentum distribution $\propto r^{1/2}$ for Keplerian disc it
takes more time to accrete matter from remote orbits,
consequently, the accretion rate will decrease slightly with time
and this fall will be displayed as the extended slope plateau.
Interestingly, since the angular momentum transfer is determined not by
the parameters of binary but by the characteristics of the wave,
this is can serve as the explanation of {\it approximately
constant} value of the plateau slope for different systems. The
value of the slope is equal to $\approx9$ days/mag, i.e. the
brightness decreases by one stellar magnitude (or, the same, the
accretion rate decreases 2.5 times) during 9 days. It means that
in 9 days the more distant ($2.5^2r_{accr}\simeq0.1A$) portion of
the disc than previous one begins to accrete. Based on these
considerations and having the measurements of superoutburst
duration and the plateau slope we can evaluate the wave size (i.e.
the distance between the initialization and termination points)
and deduce some physical characteristics of the disc. Indeed when
knowing both the wave size (i.e the radius of accreted part of the
disc) and energy release during the superoutburst we can evaluate
the mass of the disc and its density as well. The variations in
superoutburst duration is apparently due to variations of the wave
size, so for given binary parameters and, consequently, for known
maximum disc radius we can deduce the maximum superoutburst
duration for this system. Adopting $q \simeq 0.1$ for a typical
SU~UMa star we can get the maximum radius for the disc (and
maximum possible  wave size as well) is $0.54A$ [\ref{pac77}],
hence the brightness can change up to $\sim 4.25^m$ at most, and
maximum superoutburst duration is $\sim 40$ days. Adopting again
that only a half of disc mass accretes and accepting the density
of disc matter to be constant we can find that the wave size is
$\sqrt 2$ less than disc radius. So new estimation for maximum
brightening is $\sim 3.9^m$, and for maximum duration of
superoutburst is $\sim 35$ days. The analysis of observational
data shows that {\it there are no} SU~UMa type stars
\q not satisfying these estimations.

{\bf Connection between superoutbursts and normal outbursts
(peculiarities \#6 and \#7 from Section~2).} Our mechanism suggests
that superoutbursts are caused by abrupt increasing of the
accretion rate. Normal outbursts are caused by increase of the
accretion rate too. Hence, though the increase of accretion
rate in these two cases have different nature (formation of the
``precessional" wave for superoutburst and disc instability for
normal outburst) their observational manifestations can coincide
for early stages. It explains the superoutburst
beginning as a normal outburst (peculiarity~\#6). As about the
peculiarity \#7 the absence of observations of normal outbursts
occurring on, or immediately at the end of, a superoutburst
doesn't prove the connection between these types of outbursts. In
our mechanism superoutbursts and normal outbursts have different
nature but all the same we can explain the peculiarity~\#7 as
follows: superoutburst destroys the entire inner part of the disc
so no outbursts are possible until it will be filled again. The filling
rate is determined by efficiency of the outward transport of angular
momentum (for instance, due to the turbulent viscosity) but even
if the efficiency is rather high it takes a considerable
refilling time which is comparable to the outburst duration.

{\bf Appearance of superhump (peculiarity \#8 from Section 2).}
The superhump appearance is also a consequence of the
``precessional" wave formation in the disc. Increasing of accretion
rate after gas particle passing the wave is spatially localized in
azimuth direction, hence matter arrives to the accretor surface
over rather compact zone. During the outburst developing both gas
heating and difference in rotational velocities of the accretor
and the wave will increase this zone up to the forming of a belt.
Nevertheless there will be a ``kernel" of energy release
characterized by the increased values of accretion rate. This
``kernel" is rather compact and located on the accretor surface so
it can't be detectable at some orbital phases.
 An observer registers the
formation of superhump in the moment of egress of ``kernel" from
eclipse, when the ``kernel" is oriented toward the observer. The
``kernel" is connected to the ``precessional" wave so its rotational
velocity is determined by the velocity of the wave. The ``kernel"
amplitude is determined relatively to the energy release on the
entire surface so the amplitude depends on the properties of the
accretor and the disc but not on those of the wave. The fact that
the superhump is usually registered some time after the beginning
of superoutburst is a consequence of the ``kernel" compactness. If
we suppose that at the moment of superoutburst beginning the wave
location (and, therefore, the position of the superhump) is
azimuthally distributed with uniform probability then, in average,
it takes about a half of precessional period for ``kernel" to
egress from eclipse (i.e. to reach the direction toward the
observer) and will be registered as superhump.

{\bf The details of the disc structure from observations of
superhumps (peculiarities \#9 and \#10 from Section 2).}
Observations of light curves of eclipsing SU~UMa systems display
appearance of dips at phases $\sim0.2-0.25$ and $0.75-0.8$ both in
optical and UV bands. In our model we can find two arms of the
tidal shock wave on these phases. The heating of gas on shocks
results in increasing of the disc thickness and can
explain the observed dips. The peculiarity~\#10 dealing with
non-circular motion of the disc fraction with period $P_s$ also
can be explained in the framework of our model since
``precessional" spiral wave is formed in the region of
non-circular motion and rotates with the superhump period. These
regions being observed can serve as the direct confirmation of
our model.

\pagebreak

{\bf The ``late superhump" appearance (peculiarity \#11 from
Section~2).} Observations show that for some SU~UMa stars after
completion of superoutburst there are both normal orbital humps
and also a modulation of brightness at the superhump period but
shifted in phase at $\sim 180^\circ$; these are known as the ``late
superhumps". In the framework of our model the appearance of ``late
superhump" can be explained as follows: (i) the ``precessional"
spiral wave is formed in the region of apastrons of flowlines;
(ii) during the superoutburst the accretion of matter results in
the formation of an empty zone (or, more exactly, the zone of a
decreased density) in the inner part of the disc. The shape of
this zone being non-circular -- it is gaunt in the area where the
wave was and is located closer to the accretor at the opposite
side (in the regions of periastrons of flowlines); (iii) after the
end of superoutburst and the wave disappearing we will have an
elliptical ring of matter instead of the disc, the periastron of
this ring is shifted in phase at $\sim 180^\circ$ as compared to
the former location of the wave (or, the same, the superhump
phase); (iv) after the superoutburst the transport of angular
momentum and, consequently, the accretion are due to viscosity
which is the process with uniform azimuthal distribution. Hence,
gas particles will lose angular momentum in an axially-symmetrical
way, so they will reach the accretor surface faster in the region
of closer starting positions, i.e. in the region of periastron of
the elliptical ring. This results in the formation of the ``late
superhump", that is the modulation of brightness at the superhump
period but shifted in phase at $\sim 180^\circ$. The ``late
superhump" lifetime is determined by the time of circulization of
flowlines in the disc.

{\bf Superhump characteristics (peculiarities \#12--\#16 from
Section~2).} The superhump observed in every SU~UMa star for which
high speed photometry during a superoutburst has been obtained is
the evidence of the presence of a common mechanism of formation of
both superhump and superoutburst. Our model suggests that both
superhump and superoutburst are generated by appearance of
``precessional" spiral wave in the disc. This wave is formed in the
region of non-circular flowlines and the velocity of its rotation
is determined by retrograde precession of flowlines ($\sim$ few
hundredth of revolution per one orbital period), hence the
superhump peculiarities \#13 (connection of superhump and
non-circular rotation of the disc) and \#14
\q (3--7\% excess of
superhump over the orbital period)
are natural consequences of the formation mechanism. The
superhump period is determined by the period of the wave
$P_{wav}$ and orbital period $P_{orb}$ in accordance to
$$
P_s=\frac{P_{wav}P_{orb}}{P_{wav}-P_{orb}}\,.
$$
The decrease of the superhump period during the superoutburst
(peculiarity~\#15) is also quite natural if considering the
appearance of superhump to be due to the formation of the
``precessional" spiral wave in the disc. Indeed, the value of the
precessional period is the average between the period of ``fast"
outer flowlines and ``slow" inner ones (see [\ref{prec_1}] for
details). As the superoutburst develops the wave size will
decrease and ``slow" inner flowlines will exert more influence. The
wave rotation velocity will decrease (i.e., $P_{wav}$ will
increase), and it will result in diminishing of superhump period.
As mentioned above, the appearance of superhumps being independent
on the binary inclination (peculiarity \#16) is explained by the
existence of the compact ``kernel" of energy release which is
spatially localized in azimuthal direction, so the brightness
variation is due to the ``kernel" being located on the
visible/invisible side of accretor.

{\bf The superhump amplitude (peculiarities \#17 and \#18 from
Section~2).} The typical superhump amplitude is $\sim 0.3-0.4^m$,
this means the energy release in the ``kernel" is $\sim$10\% of the
energy release on the rest of accretor surface. It is quite
natural since up to the moment of first registration of superhump
by observer the accretor revolves few times, so the accretion zone
will be a belt with a small ``kernel" (see also comment on the
peculiarity~\#8). For the systems where the interregnum is absent
and superhump is registered right away after superoutburst onset
[\ref{kv}] the ``kernel" is seen just in the beginning and the belt
is formed later, as far as the matter accretes. In particular,
this effect is displayed in changing of the superhump shape during
the outburst: the initial shape is asymmetric (since the rate of
energy release as well as the size of region of energy release are
different for phases prior and after the superhump), but when the
belt is formed the superhump shape becomes symmetrical.  The
amplitude of superhumps decreasing faster than the system
brightness (peculiarity~\#18) can be explained as follows: as the
superoutburst develops the accretion occurs from more and more
high orbits, so the ``kernel" localization in azimuthal direction
becomes weaker. This results in decreasing of ratio of energy
release in the compact ``kernel" to the total energy release, i.e.
to the amplitude of superhumps decreasing faster than the
superoutburst amplitude.

{\bf Inverse correlation between the superhump
color temperature and brightness (peculiarity \#19
from Section 2).} The superhump is observed in the moment of
time when the ``kernel" of energy release zone is oriented
towards the observer. In this time there is the
density spiral wave of ``precessional" type between the accretor
ant the observer, this results in reddening due to enlarge
absorption in the superhump maximum in comparison with its
minimum when the wave doesn't prevent the observations.

{\bf Superhump light sources (peculiarity \#20 from Section~2).}
When applying the eclipse mapping technique to SU~UMa stars
[\ref{odon}] it can be found three light sources in the outer
regions of the disc located in areas coinciding with constituents
of presented model: two arms of tidal spiral shock and ``hot line"
shock. The fact that the eclipse mapping technique doesn't display
the compact energy release zone on the accretor surface is thought
to be due to the ``excess of azimuthal symmetry" of the method
[\ref{warner}], so it smears the fine azimuthal details. This
doesn't prevent the investigations of the outer parts of the disc,
but the method fails in revealing of bright spot on the accretor
surface.

Resuming the comparison of computed and observed peculiarities for
superoutbursts and superhumps  we can state that for the first
time {\it all, with no exceptions,} peculiarities (including even
``late superhumps" that are very hard to interpret) are naturally
explained in the framework of unified model.

\section*{Acknowledgements}

The work was partially supported by Russian Foundation for Basic
Research (projects NN 02-02-16088, 02-02-17642, 03-01-00311,
02-02-16622), by Science Schools Support Programme (project N
162.2003.2), by Federal Programme ``Astronomy", by Presidium RAS
Programmes ``Mathematical modelling and intellectual systems",
``Nonstationary phenomena in astronomy", and by INTAS (grant N
00-491).  OAK thanks Russian Science Support Foundation for
the financial support.

\section*{References}

\begin{enumerate}

\item
\label{prec_1}
D.V.Bisikalo, A.A.Boyarchuk, P.V.Kaygorodov,
O.A.Kuznetsov and T.Matsuda, Astron. Reports [in press].

\item
\label{warner} B.Warner, {\it Cataclysmic Variable Stars}
(Cambridge Univ. Press, Cambridge, 1995).

\item
\label{vogt74} H.C.Vogt, Astron. \& Astrophys. {\bf 36}, 369.
(1974).

\item
\label{papa78} J.Papaloizou and J.E.Pringle, Astron \&
Astrophys. {\bf 70}, L65 (1978).

\item
\label{papa79} J.Papaloizou and J.E.Pringle, Mon. Not. R.
Astron. Soc. {\bf 189}, 293 (1979).

\item
\label{vogt80} H.C.Vogt, Astron. \& Astrophys. {\bf 88}, 66
(1980).

\item
\label{osaki85} Y.Osaki, Astron. \& Astrophys. {\bf 144}, 369
(1985).

\item
\label{osaki89} Y.Osaki, Publ. Astron. Soc. Pacific {\bf 41},
1005 (1989).

\item
\label{duschl89} W.J. Duschl and M.Livio, Astron. \& Astrophys.
{\bf 241}, 153 (1989).

\item
\label{vogt82} H.C.Vogt, Astrophys. J. {\bf 252}, 563 (1982).

\item
\label{mine88} S.Mineshige, Astrophys. J. {\bf 355}, 881 (1988).

\item
\label{white88} R.Whitehurst, Mon. Not. R. Astron. Soc. {\bf
232}, 35 (1988).

\item
\label{white91} R.Whitehurst and A.King, Mon. Not. R. Astron.
Soc.  {\bf 249}, 25 (1991).

\item
\label{lubow91} S.H.Lubow, Astrophys. J. {\bf 381}, 268 (1991).

\item
\label{pac77} B.Paczy\'nski, Astrophys. J. {\bf 216}, 822 (1977).

\item
\label{osaki96} Y.Osaki, Publ. Astron. Soc. Pacific {\bf 108},
39 (1996).

\item
\label{a1} E.Meyer-Hoffmeister and H.Ritter, in ``Realm of
Interacting Binary Stars", eds J.Sahade, Y.Kondo and G.McClusey,
p.143.

\item
\label{a2} J.K.Cannizzo, in ``Accretion Disks in Compact Stellar
Systems", ed. J.C.Wheeler, p.6.

\item
\label{murray00} J.R. Murray, B.Warner and D.T.Wickramasinghe,
Mon. Not. R. Astron. Soc. {\bf 315}, 707 (2000).

\item
\label{rolfe01} D.J.Rolfe, C.A.Haswell and J.Patterson, Mon.
Not. R. Astron. Soc. {\bf 324}, 529 (2001).

\item
\label{ham} V.Buat-M\'enard and J.-M.Hameury, Astron. \&
Astrophys. {\bf 386}, 891 (2003).

\item
\label{woerd88} H. van der Woerd, M. van der Klis, J. van
Paradijs, K.Beuermann and C.Motch, Astrophys. J. {\bf 330}, 911
(1988).

\item
\label{hessman92} F.V.Hessman, K.H.Mantel, H.Barvig and
R.Shoembs, Astron. \& Astrophys., {\bf 263}, 147 (1992).

\item
\label{di12} D.V.Bisikalo, A.A.Boyarchuk, P.V.Kaygorodov and
O.A.Kuznetsov, Astron. Reports {\bf 47}, 809 (2003).

\item
\label{di8} D.V.Bisikalo, A.A.Boyarchuk, O.A.Kuznetsov and
V.M.Chechetkin, Astron. Reports {\bf 44}, 26 (2000)
[preprint astro-ph/9907087].

\item
\label{di10} D.Molteni, D.V.Bisikalo, O.A.Kuznetsov and
A.A.Boyarchuk, Mon. Not. R. Astron. Soc. {\bf 327}, 1103 (2001).

\item
\label{di11} A.A.Boyarchuk, D.V.Bisikalo, O.A.Kuznetsov and
V.M.Chechetkin, \q {\it Mass Transfer in Close Binary Stars},
(Taylor \& Francis, London, 2002).

\item
\label{Sawada86} K.Sawada, T.Matsuda and I.Hachisu, Mon.  Not.  R.
Astron. Soc. {\bf 219}, 75 (1986).

\item
\label{spiral1} K.Sawada, T.Matsuda and I.Hachisu, Mon. Not. R.
Astron. Soc. {\bf 221}, 679 (1986).

\item
\label{spiral2} K.Sawada, T.Matsuda, M.Inoue and I.Hachisu,
Mon. Not. R. Astron. Soc. {\bf 224}, 307 (1987).

\item
\label{kumar} S.Kumar, Mon. Not. R. Astron. Soc. {\bf 223}, 225
(1986).

\item
\label{kv} W.Krzeminski and N.Vogt, Astron. \& Astrophys. {\bf
144}, 124 (1985).

\item
\label{odon} D.O'Donoghue, Mon. Not. R. Astron. Soc. {\bf 246},
29 (1990).

\end{enumerate}

\end{document}